\documentclass[aps,pra,graphicx,twocolumn]{revtex4-2}

\usepackage{graphicx}% Include figure files
\usepackage{dcolumn}% align table columns on decimal point
\usepackage{bm}% bold math
\usepackage{braket}
\usepackage{amsmath}
\usepackage{amsfonts}
\usepackage{amssymb}
\usepackage{autobreak}
\usepackage{enumerate}
\usepackage{appendix}
\usepackage{booktabs}
\usepackage{dcolumn}% Align table columns on decimal point
\usepackage{bm}% bold math
\usepackage[]{threeparttable}%%%%%%%这行记得加上
\usepackage{float}
\usepackage[]{tabularx}
\usepackage[]{multirow}
\usepackage{array}
\makeatletter
\let\newfloat\newfloat@ltx
\makeatother
\usepackage{algorithm,algorithmic}
\usepackage[colorlinks,linkcolor=blue,citecolor=blue,urlcolor=blue]{hyperref}
\usepackage{color}

\newtheorem{definition}{Definition}

\usepackage{epstopdf}
\usepackage{verbatim}
\usepackage[caption=false,font=scriptsize,labelfont=sf,textfont=sf]{subfig}

\begin{document}
%\preprint{APS/123-QED}

\title{Robust hyperentanglement self testing}
\author{Yu-Hao Wang$^{1}$, Xing-Fu Wang$^{1}$, Ming-Ming Du$^{2}$, Shi-Pu Gu$^{2}$,  Wei Zhong$^{3}$}
\author{Lan Zhou$^{1}$}
\email[]{zhoul@njupt.edu.cn}
\author{Yu-Bo Sheng$^{2}$}
\affiliation{
$^1$College of Science, Nanjing University of Posts and Telecommunications, Nanjing 210023, China\\
$^2$College of Electronic and Optical Engineering, \& College of Flexible Electronics (Future Technology), Nanjing
University of Posts and Telecommunications, Nanjing, 210023, China\\
$^3$Institute of Quantum Information and Technology, Nanjing University of Posts and Telecommunications, Nanjing, 210003, China\\
}

\begin{abstract}
Hyperentanglement, which refers to entanglement encoded in two or more independent degrees of freedom (DOFs), is a valuable resource for the future high-capacity quantum network.  Certifying hyperentanglement sources working as intended is critical for the hyperentanglement-based quantum information tasks. Self testing is the strongest certification method for quantum state and measurement under minimal assumptions, even without any knowledge of the devices' inner workings. However, the existing self testing protocols all focus on one-DOF entanglement, which cannot self test the multi-DOF entanglement.
In the paper, we propose a hyperentanglement self testing framework. We take the self testing for the polarization-spatial-mode hyperentangled Bell states as an example. The self testing is based on the violation of two-dimension Clauser-Horne-Shimony-Holt (CHSH) test in each DOF independently. The two-step swap isometry circuits are proposed for self testing the entanglement in spatial-mode and polarization DOFs, respectively. All the sixteen polarization-spatial-mode hyperentangled Bell states can be self tested.
Our hyperentanglement self testing framework has three advantages. First, it is a general hyperentanglement self testing framework, and can be extended to self test multi-DOF hyperentanglement and multipartite hyperentanglement. Second, it can provide the robust hyperentanglement self testing and establish the relation between the lower bound of fidelity  and the imperfect violation of CHSH inequality in each DOF.  Third, it is feasible with current experimental technology. Our hyperentanglement self testing framework provides a promising way to certify complex hyperentanglement sources, and has potential application in future high-capacity quantum network.
\end{abstract}

\maketitle

\section{Introduction}
Entanglement serves as a fundamental resource in quantum communication \cite{QKD1,QT1,QSS,QT2,QSDC,QSDC2}, distributed quantum machine learning \cite{learning1,learning2} and quantum computation \cite{computing1,computing2}.
The entanglement simultaneously encoded in two or more DOFs is defined as the hyperentanglement \cite{hyperentanglement1,hyperentanglement2,hyperentanglement3}. Hyperentanglement has attracted increasing attention due to its abilities to enhance channel capacity \cite{capacity1,capacity2,2-14zhong2024hyperentanglement,2-15kim2021noise}, facilitate comprehensive Bell-state measurement (BSM) \cite{HBSM1,HBSM2,HBSM3,2-5li2016complete,2-3wang2019complete,one-step} and realize high-efficient entanglement purification \cite{HBSM2,purification0,purification2,purification3,purification4,purification5}. A large number of hyperentanglement generation protocols have been proposed and experimentally demonstrated, such as the polarization-spatial-mode hyperentanglement \cite{source1,source2n}, polarization-frequency
 hyperentanglement \cite{source3}, polarization-orbital angular momentum (OAM) \cite{source4,source5}, and polarization-time-bin hyperentanglement \cite{source6,2-17wang2023quantum}. Recently, the feasible generation protocols of two-photon three-DOF hyperentangled Bell state and two-DOF hyperentangled GHZ state were proposed \cite{source7,source8,2-7zhao2025direct}.

 In practical applications, certifying that hyperentanglement sources work as intended is critical and challenging. On one hand, the hyperentanglement source devices become increasingly complex \cite{certification}. On the other hand, the devices may be affected by noise and imperfections that are unknown to
 the user. Self testing, introduced by Mayers and Yao in 2004 \cite{MaYao}, is the strongest form of certifying
 the quantum state and measurements based on the input-output statistics, without requiring knowledge of the devices' internal workings \cite{1-12vsupic2020self}. Self testing only requires minimal assumptions: the no-signalling constraint on the devices, and the validity of quantum theory, so that it constitutes a form of device independent (DI) certification \cite{baccari2020device}. In the self testing framework, the violation of Bell (Clauser-Horne-Shimony-Holt (CHSH)) inequality establishes
 the existence of quantum correlations which cannot be reproduced by local models \cite{Bell,Bellnonlocality}.
 In the following years, self testing has attained great theoretical progresses. On one hand, self testing is extended from two-qubit systems to multi-qubit systems \cite{multi1,robust6,multi2,multi3,multi1n}, such as the graph state \cite{multi1n}.
 On the other hand, robust self-testing protocols have been proposed, which can infer the imperfect quantum state and estimate the bound on its fidelity \cite{robust,robust0,robust00,robust1,robust2,robust2n,robust3,robust4,robust5,robust7,1-7bancal2021self,self,robust8,multi1n}. It is theoretically proved that all pure bipartite entanglement can be self tested \cite{robust3,robust8}.
 Recently, self testing has achieved great experimental achievements \cite{exp1,exp2,exp3,exp4n}. In 2020, Tavakoli \emph{et al.} experimentally demonstrated the self testing of a targeted nonprojective measurement in the noisy scenario \cite{exp1}. In 2021, the self testing of two significant building
 blocks (a parallel configuration) of a quantum network was experimentally realized, which contributed to the certification of the larger quantum network \cite{exp2}.
  Later, Hu \emph{et al.} reported the experiment of the robust self testing in the $^{40}Ca^{+}$ ion quantum system based on non-contextuality inequalities \cite{exp3}. Recently, a robust self testing protocol based on the elegant Bell inequality was presented, which experimentally realized the self testing of the maximally entangled state and projective measurement with extremely high fidelities \cite{exp4n}.

However, existing self testing protocols all focus on one-DOF entanglement. The hyperentanglement belongs to multi-DOF entanglement, in which the entanglement in each DOF can be operated independently. All the existing self testing protocols cannot self test multi-DOF entanglement. This fact results in the inability to certify hyperentanglement sources, severely limiting their practical applications.
 In this work, we propose the first hyperentanglement self testing framework. For convenience, we take the self testing of the polarization-spatial-mode hyperentangled Bell states as an example.
 The self testing does not require the complex high-dimension Bell-like tests, but only requires to perform the two-dimension CHSH test in each DOF independently. The violation of the CHSH inequality ensures the quantum correlation between the photons in each DOF. Then, the parties construct the anticommuting observables in both DOFs, which are used to construct the swap isometry circuits. We design the two-step swap isometry circuits for self testing the entanglement in spatial-mode and polarization DOFs, respectively. All the sixteen polarization-spatial-mode hyperentangled Bell states can be self tested.
 Our protocol provides a general hyperentanglement self testing framework, which can be extended to self test the multi-DOF hyperentanglement and multipartite hyperentanglement. Moreover, it can provide robust hyperentanglement self testing. We establish the relation between the lower bound of total fidelity  and the imperfect violation of CHSH inequality in each DOF. Meanwhile, the hyperentanglement self testing protocol can be realized with current experimental technology. Our hyperentanglement self testing framework provides a simple and feasible self testing method for the complex hyperentanglement sources, and has potential application in future hyperentanglement-based high-capacity quantum network.

This paper is organized as follows.
In Sec. \ref{section1}, we explain the self testing protocol for the maximally polarization-spatial-mode hyperentangled Bell states, including the violation of the CHSH inequalities in both DOFs, the construction of the hyperentangled anti-commute relationships, and the two-step swap isometry circuits. In Sec. \ref{section2}, we extend our protocol to the robust hyperentanglement self testing. In Sec. \ref{section3}, we discuss the performance and experimental demonstration of the hyperentanglement self testing protocol. Finally, we draw a conclusion in Sec. \ref{section4}.

\section{The self testing protocol for the maximally polarization-spatial-mode  hyperentangled Bell states}\label{section1}
\subsection{The violation of the CHSH inequalities in both DOFs}
Quantum nonlocality is a necessary ingredient for self testing quantum state.
 There are sixteen kinds of polarization-spatial-mode hyperentangled Bell states, say, $\ket{\phi^{\pm}_P}\ket{\phi^{\pm}_S}$, $\ket{\phi^{\pm}_P}\ket{\psi^{\pm}_S}$, $\ket{\psi^{\pm}_P}\ket{\phi^{\pm}_S}$, and $\ket{\psi^{\pm}_P}\ket{\psi^{\pm}_S}$, where
\begin{align}\label{eq1}
    &\ket{\phi^{\pm}_P}=\frac{1}{\sqrt{2}}(\ket{hh}\pm\ket{vv}),\nonumber \\
    &\ket{\psi^{\pm}_P}=\frac{1}{\sqrt{2}}(\ket{hv}\pm\ket{vh}),\nonumber \\
    &\ket{\phi^{\pm}_S}=\frac{1}{\sqrt{2}}(\ket{a_1b_1}\pm\ket{a_2b_2}),\nonumber \\
    & \ket{\psi^{\pm}_S}=\frac{1}{\sqrt{2}}(\ket{a_1b_2}\pm\ket{a_2b_1}).
\end{align}
The subscripts "P" and "S" represent the polarization and spatial-mode DOFs, respectively. $|h\rangle$ ($|v\rangle$) represents the horizontal (vertical) polarization. $a_1$ and $a_2$ ($b_1$ and $b_2$) represent two different spatial modes in Alice's (Bob's) location.
We take the hyperentangled state $\ket{\Psi}=\ket{\phi^{+}_P}\otimes\ket{\phi^{+}_S}$ as an example. Our work inherently assumes that the hyperentanglement source generates $m$  two-photon polarization-spatial-mode hyperentangled state $\rho$ in spatial-modes $a_1$, $a_2$, $b_1$, $b_2$ ($m$ is a large number). The generated hyperentangled states follow the identical distribution (i.i.d) principle across all trials, i.e. the hyperentangled states are assumed to be exactly identical in each round, which do not depend on past measurements and generations. As each DOF in the hyperentangled system can be operated independently, it is not required to construct complex high-dimension Bell-like tests, but only requires to independently perform the two-dimension CHSH test in each DOF.

  In detail, for each photon,
Alice and Bob randomly perform one of two possible local measurements in each DOF, denoted by $A_i$ and $B_j$ ($i,j=0,1$).
In the polarization (spatial-mode) DOF, Alice's two measurement bases are denoted as
$\{ A_{0}^{P(S)}=\frac{1}{\sqrt{2}}(\sigma_{z}^{P(Sa)}+\sigma_{x}^{P(Sa)})$ and $A_{1}^{P(S)}=\frac{1}{\sqrt{2}}(\sigma_{z}^{P(Sa)}-\sigma_{x}^{P(Sa)})\}$
while Bob's two measurements bases are denoted as
$\{B_{0}^{P(S)}=\sigma_{z}^{P(Sb)}$ and $B_{1}^{P(S)}=\sigma_{x}^{P(Sb)}\}$. Here, $\sigma_z$ and $\sigma_x$ denote the Pauli operators with the forms of
 \begin{align}
\sigma_{z}^{P} &= |h\rangle\langle h| - |v\rangle\langle v|, \qquad\qquad \sigma_{x}^{P} = |h\rangle\langle v| + |v\rangle\langle h|, \nonumber \\
\sigma_{z}^{Sa} &= |a_{1}\rangle\langle a_{1}| - |a_{2}\rangle\langle a_{2}|,\qquad  \sigma_{x}^{Sa} = |a_{1}\rangle\langle a_{2}| + |a_{2}\rangle\langle a_{1}|, \nonumber\\
\sigma_{z}^{Sb} &= |b_{1}\rangle\langle b_{1}| - |b_{2}\rangle\langle b_{2}|,\qquad  \sigma_{x}^{Sb} = |b_{1}\rangle\langle b_{2}| + |b_{2}\rangle\langle b_{1}|.
\end{align}

\begin{figure}
    \centering
    \includegraphics[width=1\linewidth]{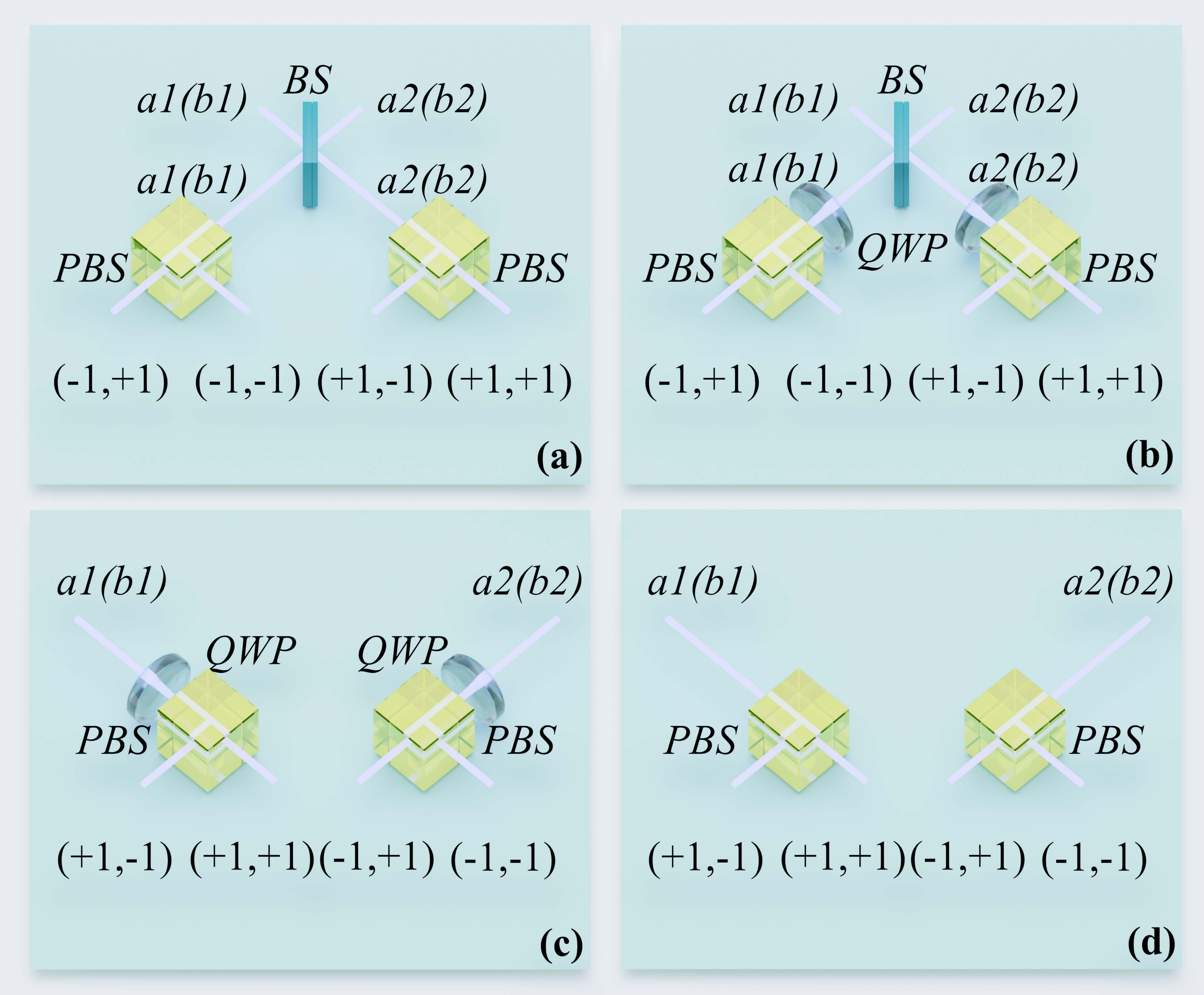}
    \caption{The linear optical apparatuses for the CHSH measurements of the polarization-spatial-mode hyperentanglement \cite{chen2003all,yang2005all}. (a) (b) (c) (d) correspond to $\sigma_{x}^S\sigma_{z}^P$, $\sigma_{x}^S\sigma_{x}^P$, $\sigma_{z}^S\sigma_{x}^P$, and $\sigma_{z}^S\sigma_{z}^P$ measurement bases, respectively.
    The polarization beam splitter (PBS)
can totally transmit the photon in $|h\rangle$ and reflect the photon in $|v\rangle$. The 50:50
beam splitter (BS) and quarter wave plate (QWP) are used to perform the Hadamard (H) operations in the spatial-mode DOF and polarization DOF, respectively.}
    \label{Fig1}
\end{figure}

The CHSH tests for the polarization-spatial-mode hyperentangled photon pairs can be realized with linear optical elements \cite{chen2003all,yang2005all} as shown in Fig. \ref{Fig1}. The measurement outcomes ($\tau$) corresponding to the measurements bases are represented by
\begin{align}\label{eq2}
    &\tau_{A}^P=\{\tau_{A_0}^P,\tau_{A_1}^P\}, \quad \tau_{B}^P=\{\tau_{B_0}^P,\tau_{B_1}^P\},\nonumber \\
    &\tau_{A}^S=\{\tau_{A_0}^S,\tau_{A_1}^S\}, \quad \tau_{B}^S=\{\tau_{B_0}^S,\tau_{B_1}^S\},
\end{align}
where $\tau_{A_i}^{P(S)},\tau_{B_j}^{P(S)}\in \{+1,-1\}$ ($i,j=0,1$). In each DOF, we can estimate the CHSH polynomial as \cite{chen2003all,yang2005all}
\begin{align}
    &I_{CHSH}^P = \langle(\tau_{A_0}^P+\tau_{A_1}^P)\tau_{B_0}^P\rangle + \langle(\tau_{A_0}^P-\tau_{A_1}^P)\tau_{B_1}^P\rangle, \nonumber\\
    &I_{CHSH}^S = \langle(\tau_{A_0}^S+\tau_{A_1}^S)\tau_{B_0}^S\rangle + \langle(\tau_{A_0}^S-\tau_{A_1}^S)\tau_{B_1}^S\rangle,
\end{align}
where  $\langle \tau_{A_i}^{P(S)}\tau_{B_j}^{P(S)}\rangle$ is defined as the probability of $\tau_{A_i}^{P(S)}=\tau_{B_j}^{P(S)}$ subtracting to that of $\tau_{A_i}^{P(S)}\neq\tau_{B_j}^{P(S)}$ ($i,j=0,1$).

 The hyperentangled state $\ket{\Psi}=|\phi^{+}_{P}\rangle\otimes |\phi^{+}_{S}\rangle$ can lead to the maximal violation of both CHSH inequalities, say, $I_{CHSH}^P = 2\sqrt{2}$ and $I_{CHSH}^S= 2\sqrt{2}$.  Similarly, if we consider the other hyperentangled states, we can also construct suitable CHSH polynomials to realize the maximal violation of the CHSH inequalities in both DOFs.

\subsection{Hyperentangled anti-commute relationships}
In the self testing framework, the target state is defined as the reference state. The practical state $\rho_{AB}$ corresponding to the actual experiment is called the physical state. To infer a particular state in the DI scenario, we need to define an equivalence between the reference state and the physical state. Mayers and Yao have proved \cite{MaYao} that when the statistics of a
 physical experiment agree with that of the reference experiment, the physical experiment is
 equivalent to the reference experiment, under a particular notion of equivalence.

 The following operations performed on the quantum state can preserve the entanglement feature of the outcome state in a DOF. In other words, any of the following operations is undiscoverable from the perspective of statistics \cite{robust00}
\begin{enumerate}[1.]
    \item Local changes of basis.
    \item Adding ancilla to physical systems, prepared in any joint state (the measurement does not act
 on them).
    \item Changing the action of the observables outside the support of the state.
    \item Locally embedding the state and operators in a larger (or smaller) Hilbert space.
\end{enumerate}

 \begin{definition}
In the one-DOF entanglement scenario, the reference (physical) experiment is described by an two-partite state $\ket{\psi}_{AB}$ ($\ket{\psi'}_{AB}$) in the Hilbert space $\mathcal{AB}$ ($\mathcal{A'B'}$) with the local measurement $M$ ($M'$). It is proved that the reference experiment is equivalent to the physical experiment if there exists local isometries  $\Phi_A$ and $\Phi_B$ with the functions of
    \begin{equation}
        \Phi_A\otimes \Phi_B: \mathcal{H_A\otimes H_B} \longrightarrow \mathcal{H_{A'A}\otimes H_{B'B}},
    \end{equation}
where $\Phi_A$ and $\Phi_B$ can be realized with local operations.
\end{definition}

Previous self testing protocols \cite{MaYao,robust0,robust00} often embed the initial state using local
 auxiliary state $|aux\rangle=|00\rangle_{A'B'}$ and perform a local
 unitary transformation as
\begin{align}
& \Phi_A\otimes\Phi_B[|\psi'\rangle_{AB}\otimes|aux\rangle_{A'B'}]\nonumber\\
 &=|junk\rangle_{AB}\otimes |\psi\rangle_{A'B'}.
 \end{align}
  $|junk\rangle$ represents a junk state after the reference state has been extracted.

 Similar with the self testing protocol for one-DOF entanglement, the central step of the hyperentanglement self testing from correlations achieving the maximal violation of the CHSH inequalities in both DOFs is to prove that Alice's and Bob's local
 observables in each DOF anticommute on the support of the hyperentangled Bell states. Once this is
 achieved, the anticommuting observables can be used to build the required local isometries.

 Here, we rewrite $\sigma_{z}^{P(S)}=Z^{P(S)}$, $\sigma_{x}^{P(S)}=X^{P(S)}$, and $D^{P(S)}=\frac{1}{\sqrt{2}}(X^{P(S)}+Z^{P(S)})$, so that $A_{0}^{P(S)}=\frac{1}{\sqrt{2}}(Z^{P(S)}+X^{P(S)})$, $A_{1}^{P(S)}=\frac{1}{\sqrt{2}}(Z^{P(S)}-X^{P(S)})$, $B_{0}^{P(S)}=Z^{P(S)}$ and $B_{1}^{P(S)}=X^{P(S)}$. We first consider the polarization DOF and obtain the relations as
\begin{align}\label{anti-commute-1}
     X_{A}^P\otimes I_{B}^P\ket{\phi^+_P}&=I_{A}^P\otimes X_{B}^P\ket{\phi^+_P}=\ket{\psi^+_P}, \nonumber \\
     Z_{A}^P\otimes I_{B}^P\ket{\phi^+_P}&=I_{A}^P\otimes Z_{B}^P\ket{\phi^+_P}=\ket{\phi^-_P}, \nonumber \\
     D_{A}^P\otimes I_{B}^P\ket{\phi^+_P}&=I_{A}^P\otimes D_{B}^P\ket{\phi^+_P},\nonumber\\
    &=\frac{1}{\sqrt{2}}(\ket{\psi^+_P}+\ket{\phi^-_P}), \nonumber \\
     X_{A}^PZ_{A}^P\otimes I_{B}^P\ket{\phi^+_P}&=I_{A}^P\otimes Z_{B}^PX_{B}^P\ket{\phi^+_P}\nonumber \\
    &=-\ket{\psi^-_P}, \nonumber \\
     Z_{A}^PX_{A}^P\otimes I_{B}^P\ket{\phi^+_P}&=I_{A}^P\otimes X_{B}^PZ_{B}^P\ket{\phi^+_P}\nonumber \\
  &=\ket{\psi^-_P},
\end{align}
 From Eq. (\ref{anti-commute-1}), we can derive
 \begin{align}\label{anti-commute-2}
    \{X_A^P,Z_A^P \}=0, \quad \{X_B^P,Z_B^P \}=0. \quad(for \ket{\phi^+_P})
\end{align}
 Similarly, Eq. (\ref{anti-commute-2}) can be also obtained when the polarized Bell state is $\ket{\phi^-_P}$, $\ket{\psi^+_P}$, or $\ket{\psi^-_P}$.

 From Eq. (\ref{anti-commute-2}), we can obtain
 \begin{align}\label{anti-commute-3}
 \{A_0^P,A_1^P\}\ket{\varphi_P}
 &=\frac{1}{2}[(Z_{A}^P+X_{A}^P)(Z_{A}^P-X_{A}^P)\nonumber \\
 &+(Z_{A}^P-X_{A}^P)(Z_{A}^P+X_{A}^P)]\ket{\varphi_P}\nonumber \\
  &=\frac{1}{2}(2Z_{A}^PZ_{A}^P-2X_{A}^PX_{A}^P)\nonumber\\
  &=(I_{A}-I_{A})\ket{\varphi_P}=0,\nonumber\\
  \{B_0^P,B_1^P\}\ket{\varphi_P}&=Z_{B}^PX_{B}^P+X_{B}^PZ_{B}^P\ket{\varphi_P}=0,
\end{align}
where $\ket{\varphi_P}$ belongs to $\{\ket{\phi^\pm_P}, \ket{\psi^\pm_P}\}$.

 In the spatial-mode DOF, the observables $X^{S}$ and $Z^{S}$ act as
\begin{align}
	&X^S_{A}\ket{a_1}=\ket{a_2},\quad X^S_{A}\ket{a_2}=\ket{a_1}, \nonumber \\
    &Z^S_{A}\ket{a_1}=\ket{a_1},\quad Z^S_{A}\ket{a_2}=-\ket{a_2}, \nonumber \\
	&X^S_{B}\ket{b_1}=\ket{b_2},\quad X^S_{B}\ket{b_2}=\ket{b_1}, \nonumber \\
    &Z^S_{B}\ket{b_1}=\ket{b_1},\quad Z^S_{B}\ket{b_2}=-\ket{b_2}.
\end{align}

 Similarly as the results in the polarization DOF, in the spatial-mode DOF, it can be easily proved that
 \begin{align}\label{anti-commute-4}
    \{X_A^S,Z_A^S \}=0, \quad \{X_B^S,Z_B^S \}=0,
\end{align}
 for the four Bell states \{$\ket{\phi^\pm_S}$, $\ket{\psi^\pm_S}$\}. As a result, we can also obtain
\begin{align}\label{anti-commute-3}
    \{A_0^S,A_1^S\}=0, \quad \{B_0^S,B_1^S\}=0,
\end{align}
for all the four Bell states $\{\ket{\phi^\pm_S}, \ket{\psi^\pm_S}\}$.

The anti-commute relationships in above two DOFs are crucial for constructing the hyper-isometries. In the self testing framework, the swap isometry circuit provides a standard method in which a user applies
 local unitary operations on auxiliary systems of known dimension to
 extract the target state from an unknown system. Here, we first introduce the construction of local isometry in one-DOF scenario from the anti-commute observables.

\begin{figure}
    \centering
    \includegraphics[width=0.9\linewidth]{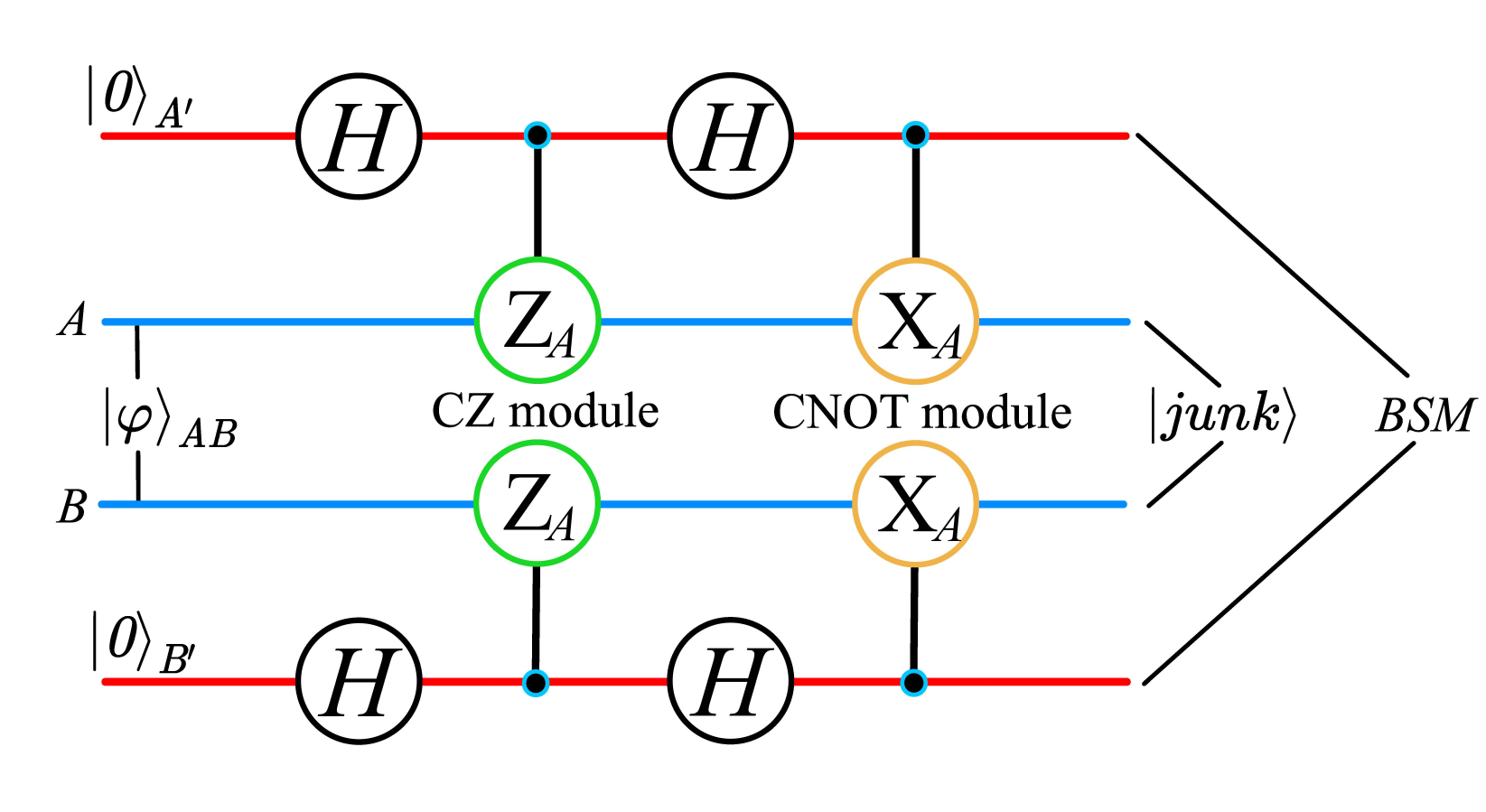}
    \caption{The theoretical swap isometry circuit in the one-DOF scenario. $H$ represents the Hadamard operation. $Z$ and $X$ represent the $\sigma_z$ and $\sigma_x$  operations in the controlled-Z (CZ) and controlled-not (CNOT) modules, respectively \cite{MaYao,robust0,robust00}.}
    \label{Fig2}
\end{figure}

The theoretical swap isometry circuit for $\Phi_A$ and $\Phi_B$ in the one-DOF scenario is shown in Fig.~\ref{Fig2} \cite{MaYao,robust0,robust00}. The physical state $\ket{\varphi}$ and the auxiliary state $|00\rangle_{A'B'}$ enter this swap isometry circuit. The whole state $|\varphi\rangle_{AB}\otimes|00\rangle_{A'B'}$ will evolve to
\begin{align}\label{Isometry-1}
    &\Phi_A\otimes \Phi_B[\ket{\varphi}_{AB}\otimes|00\rangle_{A'B'}]\nonumber\\
    &= \frac{1}{4}(I+Z_A)\otimes(I+Z_B)\ket{\varphi}\ket{00}_{A'B'} \nonumber \\
    &  +\frac{1}{4}(I+Z_A)\otimes X_B(I-Z_B)\ket{\varphi}\ket{01}_{A'B'} \nonumber \\
    & +\frac{1}{4}X_A(I-Z_A)\otimes(I+Z_B)\ket{\varphi}\ket{10}_{A'B'}\nonumber \\
    & +\frac{1}{4}X_A(I-Z_A)\otimes X_B(I-Z_B)\ket{\varphi}\ket{11}_{A'B'}.
\end{align}

Based on Eq. (\ref{anti-commute-1}) and Eq. (\ref{anti-commute-2}), the state in Eq. (\ref{Isometry-1}) with $\ket{\varphi}_{AB}=\ket{\phi^+}_{AB}$ finally transforms to
\begin{align}\label{Isometry-2}
    &\Phi_A\otimes \Phi_B[\ket{\phi^+}_{AB}\otimes|00\rangle_{A'B'}]= \ket{\phi^+}_{A'B'}\otimes\ket{00}_{AB}.
\end{align}
For the other three Bell states, after the circuit, the whole state $\ket{\varphi}_{AB}\otimes|00\rangle_{A'B'}$ can evolve to
\begin{align}\label{Isometry-3}
    &\Phi_A\otimes \Phi_B[\ket{\phi^-}_{AB}\otimes|00\rangle_{A'B'}]= \ket{\phi^-}_{A'B'}\otimes\ket{00}_{AB}, \nonumber\\
    &\Phi_A\otimes \Phi_B[\ket{\psi^+}_{AB}\otimes|00\rangle_{A'B'}]= \ket{\psi^+}_{A'B'}\otimes\ket{11}_{AB}, \nonumber\\
    &\Phi_A\otimes \Phi_B[\ket{\psi^-}_{AB}\otimes|00\rangle_{A'B'}]= -\ket{\psi^-}_{A'B'}\otimes\ket{11}_{AB}.
\end{align}
As a result, based on the BSM results on the auxiliary state in A'B' modes, one can certify the target state.

\subsection{Two-step swap isometry circuits}
 Suppose that the physical hyperentangled state is $|\varphi_S\rangle|\varphi_P\rangle$. The swap isometry circuits can be divided into two steps as follows. In theory, the parties can randomly choose half of the hyperentangled states to implement the swap isometry circuit of Step 1 and the other half of the hyperentangled states to implement the swap isometry circuit of Step 2.

\begin{figure}
    \centering
    \includegraphics[width=1\linewidth]{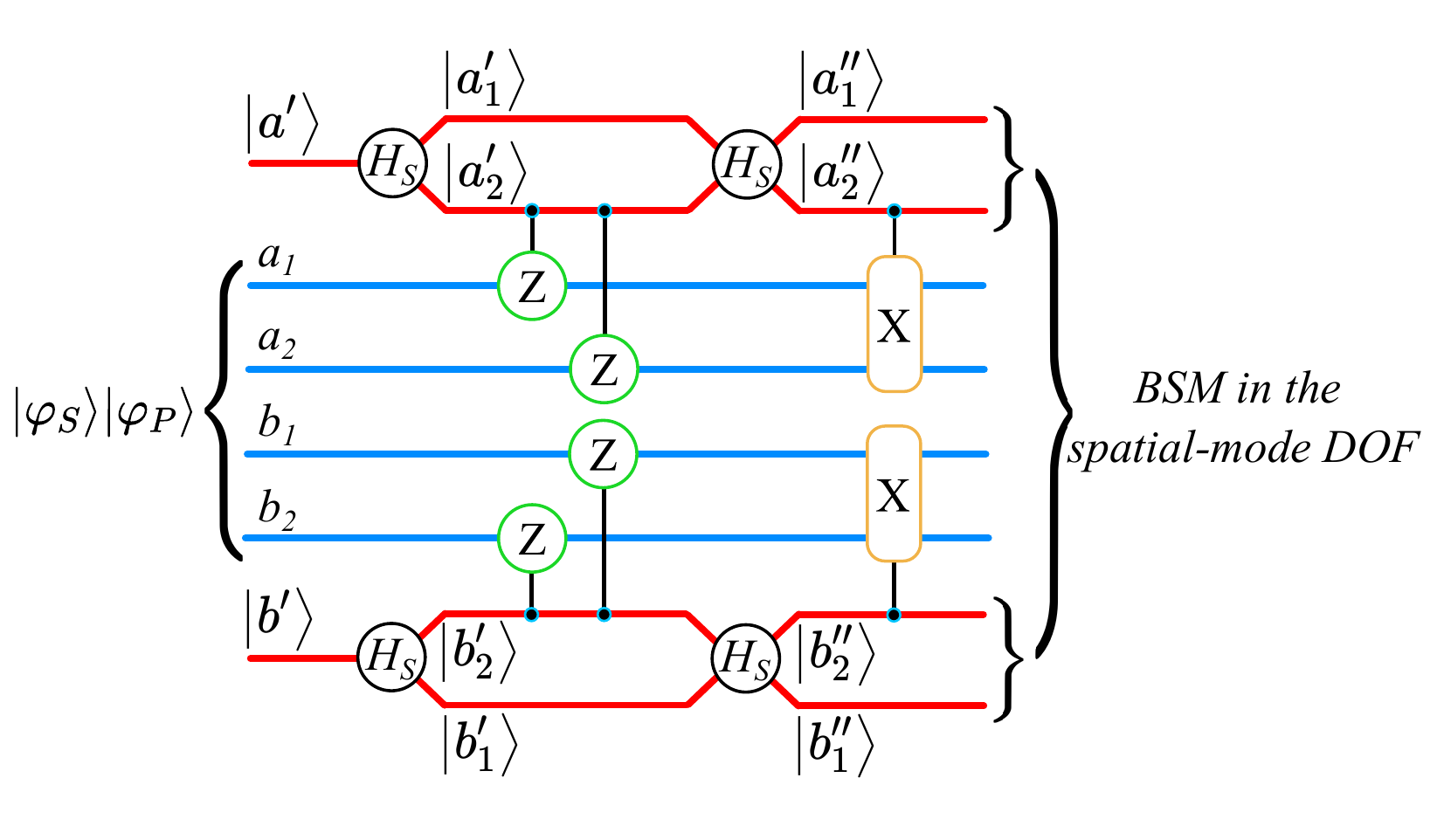}
    \caption{The swap isometry circuit in the spatial-mode DOF. Two auxiliary photons are required in the spatial modes a' and b'.  $H_S$ represents the Hadamard operation in the spatial-mode DOF.  After all the operations, the parties perform the BSM in spatial-mode DOF on the auxiliary photons in $a_1''b_1''a_2''b_2''$ modes, respectively.}
    \label{spatial}
\end{figure}

 \textbf{Step 1 The  swap isometry circuit in the spatial-mode DOF.} As shown in Fig. \ref{spatial}, the practical hyperentangled photons are in the spatial modes $a_1$, $b_1$, $a_2$, and $b_2$. Two auxiliary single photon sources generate two single photons in the spatial modes $a'$ and $b'$, respectively, where $a'$ belongs to Alice's location and $b'$ belongs to Bob's location. The physical hyperentangled photons and the auxiliary photons pass through the swap isometry circuit. The $H_S$ represents  the Hadamard operation in the spatial-mode DOF. The first column of $H_{S}s$ lead to $|a'\rangle\rightarrow\frac{1}{\sqrt{2}}(|a_1'\rangle+|a_2'\rangle)$ and  $|b'\rangle\rightarrow\frac{1}{\sqrt{2}}(|b_1'\rangle+|b_2'\rangle)$, while the second column of $H_{S}s$ lead to $|a_1'\rangle\rightarrow \frac{1}{\sqrt{2}}(|a_1''\rangle+|a_2''\rangle)$, $|a_2'\rangle\rightarrow \frac{1}{\sqrt{2}}(|a_1''\rangle-|a_2''\rangle)$, $|b_1'\rangle\rightarrow \frac{1}{\sqrt{2}}(|b_1''\rangle+|b_2''\rangle)$, and $|b_2'\rangle\rightarrow \frac{1}{\sqrt{2}}(|b_1''\rangle-|b_2''\rangle)$, respectively. The photons in $|a_2'\rangle$ ($|b_2'\rangle$) and $|a_2''\rangle$ ($|b_2''\rangle$) would cause the CZ and CNOT operations on the photons in $a_1$ ($b_1$) and $a_2$ ($b_2$) modes. After all the operations, the parties perform the spatial-mode  BSM  on the auxiliary photons in the spatial modes $a_1''$, $a_2''$, $b_1''$ and $b_2''$. Based on the anti-commute relationships in spatial-mode DOF, we can obtain
 \begin{align}\label{Isometry-2}
    &\Phi_{AS}\otimes \Phi_{BS}[\ket{\varphi_P}\otimes\ket{\phi^+_S}_{AB}\otimes|a'b'\rangle]\nonumber\\
    &=\ket{\phi^+_S}_{A'B'}\otimes\ket{\varphi_P}\otimes\ket{a_1b_1},\nonumber\\
    &\Phi_{AS}\otimes \Phi_{BS}[\ket{\varphi_P}\otimes\ket{\phi^-_S}_{AB}\otimes|a'b'\rangle]\nonumber\\
    &=\ket{\phi^-_S}_{A'B'}\otimes\ket{\varphi_P}\otimes\ket{a_1b_1}, \nonumber\\
    &\Phi_{AS}\otimes \Phi_{BS}[\ket{\varphi_P}\otimes\ket{\psi^+_S}_{AB}\otimes|a'b'\rangle]\nonumber\\
    &=\ket{\psi^+_S}_{A'B'}\otimes\ket{\varphi_P}\otimes\ket{a_2b_2}, \nonumber\\
    &\Phi_{AS}\otimes \Phi_{BS}[\ket{\varphi_P}\otimes\ket{\psi^-_S}_{AB}\otimes|a'b'\rangle]\nonumber\\
    &=-\ket{\psi^-_S}_{A'B'}\otimes\ket{\varphi_P}\otimes\ket{a_2b_2},
\end{align}
 where the four spatial-mode Bell states can be written as
 \begin{eqnarray}
 \ket{\phi^\pm_S}_{A'B'}&=\frac{1}{\sqrt{2}}(|a_1''b_1''\rangle\pm|a_2''b_2''\rangle),\nonumber\\
 \ket{\psi^\pm_S}_{A'B'}&=\frac{1}{\sqrt{2}}(|a_1''b_2''\rangle\pm|a_2''b_1''\rangle).
 \end{eqnarray}

 Based on the spatial-mode BSM results on the auxiliary photon, all the four spatial-mode Bell states can be certified. It is noted that all the above operations do not influence the polarization of the physical hyperentangled state, so that the junk state still preserve the original polarization entanglement.

 \begin{figure}
    \centering
    \includegraphics[width=0.9\linewidth]{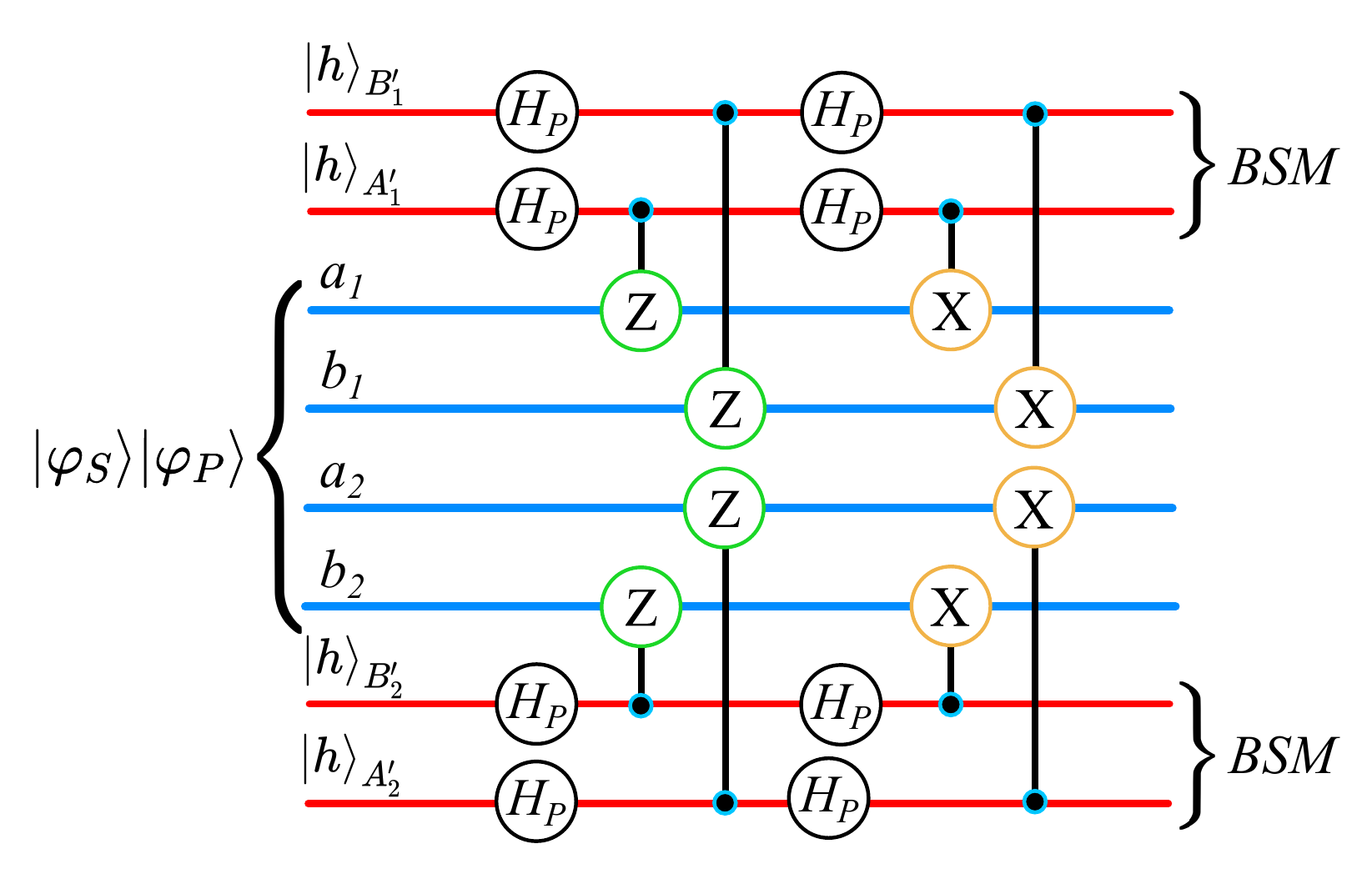}
    \caption{The swap isometry circuit in the polarization DOF. The physical hyperentangled state and two auxiliary photons pass through this swap isometry circuit. $H_P$ represents the Hadamard operation in the polarization DOF. After all the operations, the parties perform the BSMs in the polarization DOF on the auxiliary photons in the spatial modes $A_1'B_1'$ and $A_2'B_2'$, respectively.}
    \label{polarization}
\end{figure}

 \textbf{Step 2 The swap isometry circuit in the polarization DOF.} The basic principle of the swap isometry circuit in the polarization DOF is shown in Fig. \ref{polarization}. The auxiliary single photon sources generate four single photons in $|h\rangle$ in the spatial modes $A_1'$, $A_2'$, $B_1'$, and $B_2'$, respectively. Here, $A_1'$ and $A_2'$ belong to Alice's location, while $B_1'$ and $B_2'$ belong to Bob's location. The physical hyperentangled states in $a_1$, $b_1$, $a_2$, $b_2$ combined with the auxiliary states pass through the swap isometry circuit. The $H_P$ realizes the Hadamard operation as $|h\rangle\rightarrow\frac{1}{\sqrt{2}}(|h\rangle+|v\rangle)$ and $|v\rangle\rightarrow\frac{1}{\sqrt{2}}(h\rangle-|v\rangle)$. Then, the auxiliary photons and physical photons pass through the CZ and CNOT modules in the polarization DOF, successively.
 The sixteen hyperentangled Bell states can be divided into two scenarios. In the first scenario, the physical photons are in $\ket{\phi^\pm_S}$ in the spatial-mode DOF, including $\ket{\phi^\pm_P}\otimes\ket{\phi^\pm_S}$ and $\ket{\psi^\pm_P}\otimes\ket{\phi^\pm_S}$. After all the local operations, we perform the polarization BSM on the photons in $A_1'B_1'$ and $A_2'B_2'$ modes, respectively.
 Different with the entanglement in one DOF, the two photons have uncertainty in the spatial locations. As a result, it is impossible for one to deterministically extract the target state in certain output spatial mode from auxiliary systems by local unitary operations.

 We take the case that the polarized Bell state is in $a_1b_1$ as an example. The evolution process of the state in $a_1b_1$ and $A_1'B_1'$ modes is shown in Eq. (\ref{Isometry-2}) and Eq. (\ref{Isometry-3}), respectively, while the state in $a_2b_2$ and $A_2'B_2'$ modes evolves to
\begin{eqnarray}\label{Isometry-4}
&\Phi_{a_1}\otimes \Phi_{b_1}\otimes\Phi_{a_2}\otimes \Phi_{b_2}[\ket{vac}_{a_2b_2}\otimes|hh\rangle_{A_2'B_2'}]\nonumber\\
&=\ket{vac}_{a_2b_2}\otimes\frac{1}{\sqrt{2}}(|\phi^{+}_P\rangle+|\phi^{-}_P\rangle)_{A_2'B_2'}.
\end{eqnarray}

 As a result, after the circuit, the above eight hyperentangled states combined with the auxiliary states evolve to
\begin{eqnarray}\label{phy+phy+}
   & \Phi_{a_1}\otimes \Phi_{b_1}\otimes\Phi_{a_2}\otimes \Phi_{b_2}[\ket{\phi^+_P}\otimes\ket{\phi^\pm_S}\otimes|hhhh\rangle_{A_1'A_2'B_1'B_2'}]\nonumber\\
    &\rightarrow
        \ket{\phi^+_{A_1'B_1'}}\Big(\ket{\phi^+_{A_2'B_2'}}+\ket{\phi^-_{A_2'B_2'}}\Big)\ket{hh}_{a_1b_1} \nonumber \\
     &   \pm\Big(\ket{\phi^+_{A_1'B_1'}}+\ket{\phi^-_{A_1'B_1'}}\Big)\ket{\phi^+_{A_2'B_2'}}\ket{hh}_{a_2b_2},
\end{eqnarray}

\begin{eqnarray}\label{hyper-situation-2}
    & \Phi_{a_1}\otimes \Phi_{b_1}\otimes\Phi_{a_2}\otimes \Phi_{b_2}[\ket{\phi^-_P}\ket{\phi^\pm_S}\otimes|hhhh\rangle_{A_1'A_2'B_1'B_2'}]\nonumber\\
    &\rightarrow
        \ket{\phi^-_{A_1'B_1'}}\Big(\ket{\phi^+_{A_2'B_2'}}+\ket{\phi^-_{A_2'B_2'}}\Big)\ket{hh}_{a_1b_1} \nonumber \\
     &   \pm\Big(\ket{\phi^+_{A_1'B_1'}}+\ket{\phi^-_{A_1'B_1'}}\Big)\ket{\phi^-_{A_2'B_2'}}\ket{hh}_{a_2b_2},
\end{eqnarray}

\begin{eqnarray}\label{hyper-situation-3}
    & \Phi_{a_1}\otimes \Phi_{b_1}\otimes\Phi_{a_2}\otimes \Phi_{b_2}[\ket{\psi^+_P}\ket{\phi^\pm_S}\otimes|hhhh\rangle_{A_1'A_2'B_1'B_2'}]\nonumber\\
    &\rightarrow
        \ket{\psi^+_{A_1'B_1'}}\Big(\ket{\phi^+_{A_2'B_2'}}+\ket{\phi^-_{A_2'B_2'}}\Big)\ket{vv}_{a_1b_1} \nonumber \\
      &  \pm\Big(\ket{\phi^+_{A_1'B_1'}}+\ket{\phi^-_{A_1'B_1'}}\Big)\ket{\psi^+_{A_2'B_2'}}\ket{vv}_{a_2b_2},
\end{eqnarray}

\begin{eqnarray}\label{hyper-situation-4}
    & \Phi_{a_1}\otimes \Phi_{b_1}\otimes\Phi_{a_2}\otimes \Phi_{b_2}[\ket{\psi^-_P}\ket{\phi^\pm_S}\otimes|hhhh\rangle_{A_1'A_2'B_1'B_2'}]\nonumber\\
    &\rightarrow
        \ket{\psi^-_{A_1'B_1'}}\Big(\ket{\phi^+_{A_2'B_2'}}+\ket{\phi^-_{A_2'B_2'}}\Big)\ket{vv}_{a_1b_1} \nonumber \\
       & \pm\Big(\ket{\phi^+_{A_1'B_1'}}+\ket{\phi^-_{A_1'B_1'}}\Big)\ket{\psi^-_{A_2'B_2'}}\ket{vv}_{a_2b_2}.
\end{eqnarray}

From Eq. (\ref{phy+phy+}) and Eq. (\ref{hyper-situation-2}), we define the successful response events for self testing $\ket{\phi^+_P}\otimes\ket{\phi^\pm_S}$ and $\ket{\phi^-_P}\otimes\ket{\phi^\pm_S}$  as $\ket{\phi^+_{A_1'B_1'}}\ket{\phi^+_{A_2'B_2'}}$ and $\ket{\phi^-_{A_1'B_1'}}\ket{\phi^-_{A_2'B_2'}}$, respectively. It can be found that the success probability to distinguish each of the four hyperentangled Bell states $\ket{\phi^\pm_P}\otimes\ket{\phi^\pm_S}$
 is only $50\%$. However, the junk state still has entanglement in the spatial-mode DOF, which is identical with the original spatial-mode entanglement. On the other hand, all the four items in Eq. (\ref{hyper-situation-3}) and Eq. (\ref{hyper-situation-4}) can be treated as the successful response events for self testing $\ket{\psi^+_P}\ket{\phi^\pm_S}$ and $\ket{\psi^-_P}\ket{\phi^\pm_S}$, respectively. As a result, the success probability to distinguish each of the four hyperentangled Bell states $\ket{\psi^+_P}\ket{\phi^\pm_S}$ and $\ket{\psi^-_P}\ket{\phi^\pm_S}$ is 100\%. However, the junk states do not have entanglement in any DOF.

In the second scenario, the physical photons are in $\ket{\psi^\pm_S}$ in the spatial-mode DOF, including $\ket{\phi^\pm_P}\otimes\ket{\psi^\pm_S}$ and $\ket{\psi^\pm_P}\otimes\ket{\psi^\pm_S}$. Alice or Bob should perform the $\sigma_{x}^{S}$ operation on the photon in $a_1$ or $b_1$ mode with the beam displacer (BD), which can transform  $\ket{\psi^\pm_S}$ to $\ket{\phi^\pm_S}$. In this way, the eight hyperentangled Bell states can be transformed to those in the first scenario and pass through the circuit in Fig. \ref{polarization} to complete the self testing. Finally, the $\sigma_{x}^{S}$ operation should be performed to recover the original hyperentangled Bell states. In this way, the four hyperentangled states  $\ket{\phi^\pm_P}\otimes\ket{\psi^\pm_S}$ can be self tested with the success probability of 50\%, and the spatial-mode entanglement in the original practical state can be preserved in the junk state. The four hyperentangled states $\ket{\psi^\pm_P}\otimes\ket{\psi^\pm_S}$ can be self tested with the probability of 100\%, and no entanglement exists in the junk states.

\section{The robust self-testing protocol for the imperfect polarization-spatial-mode hyperentangled Bell states}\label{section2}
In Sec. \ref{section1}, we consider the self testing of the maximally hyperentangled Bell states, which can achieve the maximal violation of the CHSH inequalities in both DOFs. However, in practical scenario, it is impossible to realize the maximal violation of CHSH inequality in any DOF. On one hand, various experimental
 noise and imperfections may reduce the exact reference correlations. On the other hand, in practical demonstration, the CHSH test in each DOF only works with a finite sample size, so that the precise probabilities can be only estimated up to some statistical confidence level. The problem of estimating the fidelity (norm difference) between a practical physical bipartite state and the two-qubit maximally entangled
 state from the violation of the CHSH inequality in the one-DOF scenario has been researched since 2009 \cite{robust,robust0}. Inspired by the robust self testing protocols \cite{robust,robust0} in the one-DOF scenario, we design the robust self testing protocol for the polarization-spatial-mode hyperentangled Bell states.

We take the target hyperentangled Bell state $\ket{\phi^+_P}\otimes\ket{\phi^+_S}$ as an example. In the practical hyperentanglement system $\ket{\varphi}=\ket{\varphi_P}\otimes\ket{\varphi_S}$, the CHSH tests in two DOFs are totally independent. We assume the suboptimal violation of the spatial and polarization CHSH inequalities as
\begin{eqnarray}
I_{CHSH}^P &=& \langle(\tau_{A_0}^P+\tau_{A_1}^P)\tau_{B_0}^P\rangle + \langle(\tau_{A_0}^P-\tau_{A_1}^P)\tau_{B_1}^P\rangle=2\sqrt{2}-\epsilon_P,\nonumber\\
I_{CHSH}^S &=& \langle(\tau_{A_0}^S+\tau_{A_1}^S)\tau_{B_0}^S \rangle+ \langle(\tau_{A_0}^S-\tau_{A_1}^S)\tau_{B_1}^S\rangle=2\sqrt{2}-\epsilon_{S},\nonumber\\
\end{eqnarray}
where the parameters $\epsilon_{P}$ and $\epsilon_{S}$ are both close to zero.

In this case, the observables $X_A$, $Z_A$, $X_B$, $Z_B$ in both spatial and polarization DOFs remain close to ideal. In the polarization (spatial-mode) DOF, we can obtain
\begin{align}\label{anti2}
&|| (X_A^{P(S)}Z_A^{P(S)}+Z_A^{P(S)}X_A^{P(S)})\ket{\varphi_{P(S)}} || \le 2\varepsilon_1^{P(S)},  \nonumber\\
&|| (X_B^{P(S)}Z_B^{P(S)}+Z_B^{P(S)}X_B^{P(S)})\ket{\varphi_{P(S)}} || \le 2\varepsilon_1^{P(S)},  \nonumber\\
&|| (X_A^{P(S)}-X_B^{P(S)})\ket{\varphi_{P(S)}} || \le 2\varepsilon_2^{P(S)},  \nonumber\\
&|| (Z_A^{P(S)}-Z_B^{P(S)})\ket{\varphi_{P(S)}} || \le 2\varepsilon_2^{P(S)},
\end{align}
with $\varepsilon_1^P=2(\epsilon_P\sqrt{2})^{\frac{1}{2}}$, $\varepsilon_1^S=2(\epsilon_S\sqrt{2})^{\frac{1}{2}}$,
 $\varepsilon_2^P=4(\epsilon_P\sqrt{2})^{\frac{1}{4}}$, and $\varepsilon_2^S=4(\epsilon_S\sqrt{2})^{\frac{1}{4}}$ \cite{robust0}.

Based on the anti-commute relationships in Eq. (\ref{anti2}),  there exists a local swap isometry $\Phi_{AS}\otimes \Phi_{BS}$ in Step 1 with the function of
\begin{align}
&||\Phi_{AS}\Phi_{BS}[\ket{\varphi_P}\otimes\ket{\varphi_S}\ket{a'b'}]
-\ket{\phi^+_S}_{A'B'}\ket{\varphi_P}
\ket{junk}||\nonumber\\
& \le 3\varepsilon_1^S+5\varepsilon_2^S.
\end{align}

In Step 2, two polarization swap isometries are located in $a_1b_1$ and $a_2b_2$ spatial modes, respectively. The input photons are randomly in the spatial modes $a_1b_1$ and $a_2b_2$. On one hand, if the input state of a polarization swap isometry, such as the one in $a_2b_2$ is $|vac\rangle$, the output state in the corresponding auxiliary spatial modes $C'D'$ is  $\frac{1}{\sqrt{2}}(|\phi^{+}_P\rangle+|\phi^{-}_P\rangle)_{C'D'}$. On the other hand, if the input state of a polarization swap isometry, such as the one in $a_1b_1$ is $\ket{\varphi_P}$, the transformation process can be written as
\begin{align}
&||\Phi_{AP}\Phi_{BP}[\ket{\varphi_P}_{a_1b_1}\ket{hh}_{A'B'}]-\ket{\phi^+_P}_{A'B'}\ket{junk}_{a_1b_1}||\nonumber\\
& \le 3\varepsilon_1^P+5\varepsilon_2^P.
\end{align}
In this way, after the swap isometry circuit, the whole state evolves to
\begin{align}
&||\Phi_{AP}\Phi_{BP}[\frac{1}{\sqrt{2}}(|\varphi_P\rangle_{a_1b_1}+|\varphi_P\rangle_{a_2b_2})\otimes\ket{hhhh}_{A'B'C'D'}]\nonumber\\
&-\ket{\phi^+_{A'B'}}\Big(\ket{\phi^+_{C'D'}}+\ket{\phi^-_{C'D'}}\Big)\ket{hh}_{a_1b_1} \nonumber \\
     &   +\Big(\ket{\phi^+_{A'B'}}+\ket{\phi^-_{A'B'}}\Big)\ket{\phi^+_{C'D'}}\ket{hh}_{a_2b_2}||\nonumber \\
     &\le 3\varepsilon_1^P+5\varepsilon_2^P
\end{align}
As a result, the fidelities of the polarization and spatial-mode DOFs can be lower bounded by \cite{robust0}
\begin{align}
F_P &\ge 1-\frac{1}{4}(9\sqrt{2}\epsilon_P+2^{\frac{1}{4}}100\epsilon_P^{\frac{1}{2}}+2^{\frac{3}{8}}60\epsilon_P^{\frac{3}{4}}),\nonumber \\
F_S &\ge 1-\frac{1}{4}(9\sqrt{2}\epsilon_S+2^{\frac{1}{4}}100\epsilon_S^{\frac{1}{2}}+2^{\frac{3}{8}}60\epsilon_S^{\frac{3}{4}}).
\end{align}
We can obtain the lower bound of the total fidelity of the hyperentangled Bell state as
\begin{align}
F_t =F_P\times F_S\geq[1-\frac{1}{4}(9\sqrt{2}\epsilon_P+2^{\frac{1}{4}}100\epsilon_P^{\frac{1}{2}}+2^{\frac{3}{8}}60\epsilon_P^{\frac{3}{4}})]\nonumber \\
\times[1-\frac{1}{4}(9\sqrt{2}\epsilon_S+2^{\frac{1}{4}}100\epsilon_S^{\frac{1}{2}}+2^{\frac{3}{8}}60\epsilon_S^{\frac{3}{4}})].
\end{align}

 \begin{figure}
    \centering
    \includegraphics[width=0.8\linewidth]{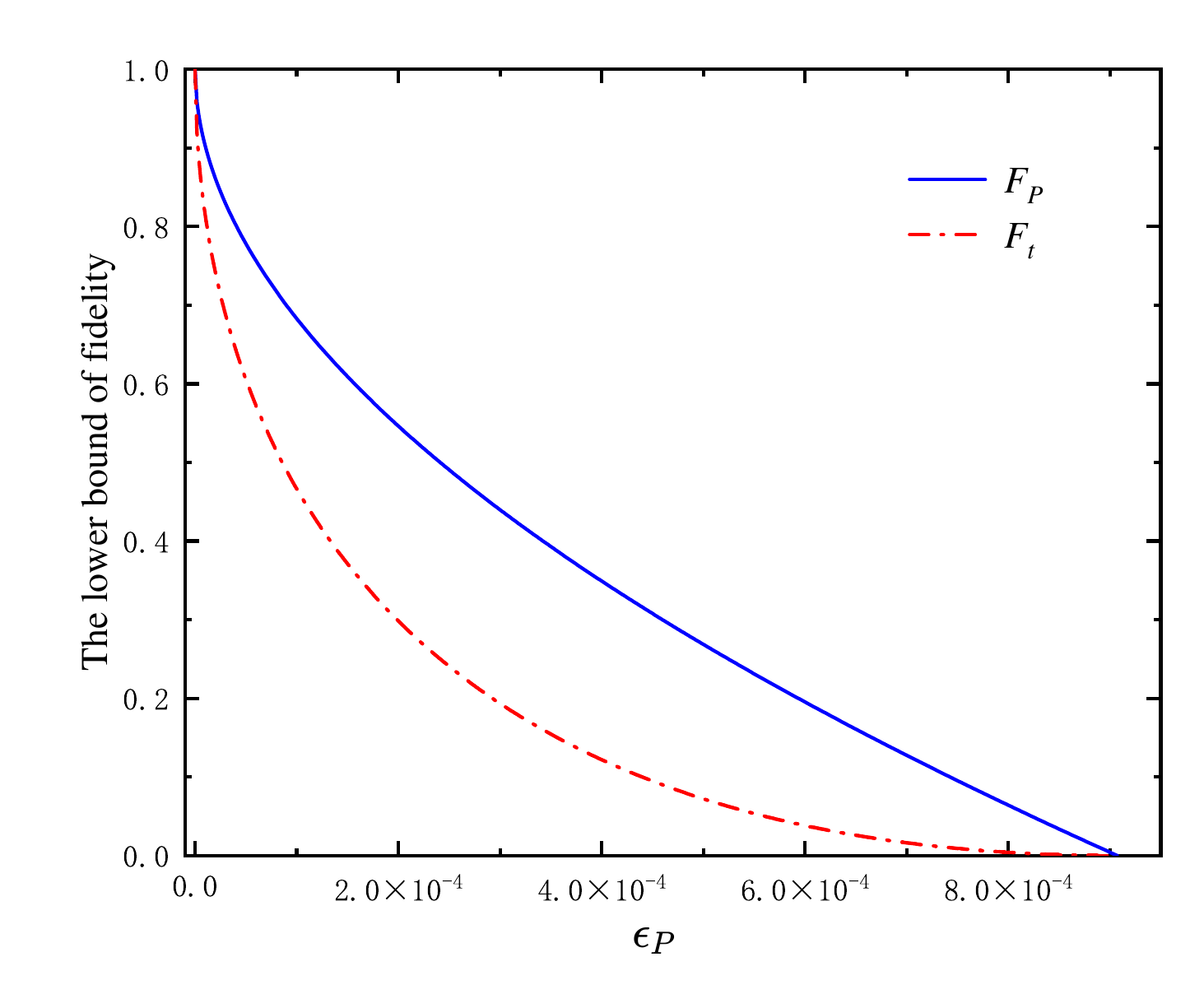}
    \caption{The lower bounds of the fidelity $F_P$ and $F_t$ as a function of the parameter $\epsilon_P$ with $\epsilon_P=\epsilon_S$.}
    \label{fidelity_simulate}
\end{figure}

 Here, we assume that $\epsilon_P=\epsilon_S$, and simulate the lower bound of $F_P$ and $F_t$ as a function of the parameter $\epsilon_P$  in Fig. \ref{fidelity_simulate}.  It can be found that the lower bounds of $F_P$ and $F_t$ are quite loose. Both the lower bounds drop rapidly with the growth of $\epsilon_P$. When $\epsilon_P\approx 2.40\times 10^{-4}$, the lower bound of $F_P$ drops to 0.5, and that of $F_t$ is drops to only 0.25. When $\epsilon_P$ increases to $\approx 9.07\times 10^{-4}$, the lower bounds of $F_P$ and $F_t$ drop to 0.

\section{Discussion}\label{section3}
We propose the self testing protocol for the polarization-spatial-mode hyperentangled Bell states. As the entanglement in both DOFs are independent, we do not require to construct the complex high-dimensional Bell-like tests, but only independently perform the two-dimension CHSH tests in both DOFs. Then, we prove that Alice's and Bob's local observables in each DOF anticommute on the Bell states, and use the anticommuting observables to construct the required swap isometry circuits in both DOFs. The swap isometry circuits can be divided into two steps. Step 1 is the spatial-mode swap isometry circuit. This swap isometry circuit can deterministically extract the reference spatial-mode Bell state from the auxiliary state by performing the local operations, while preserve the polarization feature of the physical state. As a result, the quantum state in spatial-mode can be deterministically self tested based on the BSM result of the auxiliary photons. In practical implement, the parties can randomly choose half of the practical states for Step 1 to certify the spatial-mode Bell state. After that, the remained half of physical states pass through the polarization swap isometry circuits in Step 2 to self test the polarization Bell state. In detail, as the spatial modes of the polarization Bell state is uncertain, two identical polarization swap isometry circuits should be located in both $a_1b_1$ and $a_2b_2$ modes.
Based on the spatial-mode Bell state and the BSM results of the auxiliary photon pairs in two polarization swap isometry circuits, the eight hyperentangled Bell states $\ket{\phi^\pm_P}\otimes\ket{\phi^\pm_S}$ and $\ket{\phi^\pm_P}\otimes\ket{\psi^\pm_S}$ can be self tested with the success probability of 50\%, and the junk states can preserve the original spatial-mode entanglement, while the eight hyperentangled Bell states $\ket{\psi^\pm_P}\otimes\ket{\phi^\pm_S}$ and $\ket{\psi^\pm_P}\otimes\ket{\psi^\pm_S}$ can be self tested with the success probability of 100\%, and the junk states do not have spatial-mode entanglement.

 Our protocol provides a general hyperentanglement self testing framework, which can be extended to self test the hyperentangled Bell states in other two DOFs and multi-DOF hyperentanglement (such as the polarization-spatial-mode-time-bin hyperentangled Bell states) in theory.
 Moreover, our self testing framework can also be extended to self test the multi-particle hyperentangled state, like the hyperentangled GHZ states. For example, for self testing the three-photon polarization-spatial-mode hyperentangled GHZ state with the form of $\frac{1}{\sqrt{2}}(|hhh\rangle+|vvv\rangle)\otimes\frac{1}{\sqrt{2}}(|a_1b_1c_1\rangle+|a_2b_2c_2\rangle)$, three parties first independently perform the Svetlichny tests \cite{S1,S2} in both DOFs. Then, based on the swap isometry circuit in the one-DOF GHZ state \cite{multi3}, they require to construct the two-step swap isometry circuits for the GHZ state in the spatial-mode DOF and polarization DOF, respectively. The detailed self-testing protocol will be investigated in our later work.

 In Sec. \ref{section2}, we adopt the robust self testing model from Ref. \cite{robust0} in the our hyperentanglement self testing framework to design the robust hyperentanglement self testing protocol. We provide the lower bounds for the fidelities $F_P$ and $F_S$ altered with the parameters $\epsilon_{P}$ and $\epsilon_{S}$. However, these lower bounds are quite loose, and thus the hyperentanglement self testing protocol can only be applied when the experimental statistics extremely match the ideal statistics. For enhancing the practicality of hyperentanglement self testing, we can also adopt other robust self testing models in our hyperentanglement self testing framework. For example, in 2016, Kaniewski proposed a new technique for proving
 the self testing bounds of practically relevant robustness. An improved bounds for self testing the singlet is obtained as long as the CHSH inequality violation exceeds
  about 2.11 \cite{robust2n}. In 2020, Baccari \emph{et al.} proposed a general construction of Bell inequalities for multiqubit graph states. They also adopted the method in Ref. \cite{robust2n} to estimate the fidelity altered with the relative observed violation of the corresponding violation of Bell inequalities \cite{multi1n}. Recently, a new robust self testing protocol with the violation of the elegant Bell inequality was proposed and experimentally demonstrated \cite{exp4n}. That self testing protocol allows the fidelity estimation of both quantum states and measurements for any observed violation of the elegant Bell inequality. By adopting the robust self testing protocols in one-DOF scenarios in our hyperentanglement self testing framework, it is possible to increase its robustness to noise and other imperfect factors.

  Finally, we discuss the experimental demonstration of this hyperentanglement self testing protocol. The sources for the polarization-spatial-mode hyperentangled Bell state are shown in Ref. \cite{source1,source2n}. For demonstrating the hyperentanglement self testing, we can first perform the two-dimension CHSH tests independently in both DOFs. As shown in Fig. \ref{Fig1}, the independent CHSH tests can be realized with linear optical elements PBS, QWP, and BS \cite{chen2003all,yang2005all}, which are feasible under current experimental technologies. Based on the practical violation of the CHSH inequalities in both DOFs, the lower bounds of the fidelities in both DOFs can be estimated. Then, the practical fidelities in both DOFs can be measured through the full hyperentanglement quantum state tomography, which have been experimentally realized  \cite{purification2,purification5}. We can demonstrate the hyperentanglement self testing if the practical fidelities agree with the theoretical prediction of the fidelity lower bounds. In this way, our hyperentanglement self testing is feasible under current experimental conditions.

\section{Conclusion}\label{section4}
In conclusion, hyperentanglement represents the entanglement simultaneously encoded in two or more DOFs. Hyperentanglement has high capacity, and the entanglement in each DOF can be operated independently. Benefitting to these features, hyperentanglement has become an indispensable resource for the high-capacity quantum network in the future. The certification of hyperentangled sources working as intended is critical for hyperentanglement-based quantum information tasks.
   Self testing is the strongest certification method which allows one to characterize quantum states or measurement under minimal assumptions. Existing self testing protocols all focus on one-DOF entanglement, which cannot self test the entanglement in multiple DOFs simultaneously, which largely limit the certification and practical application of hyperentanglement sources.
   In the paper, we propose the hyperentanglement self testing framework. For convenience, we take the self testing of the polarization-spatial-mode hyperentangled Bell states for example.
  Suppose two parties share a large number of identical hyperentangled states from the hyperentanglement source.
   As the entanglement in each DOF is independent, the hyperentanglement self testing does not require the complex high-dimension Bell-like tests, but only requires to perform the two-dimension CHSH test in each DOF independently. Then, we prove that two parties' local observables in each DOF anticommute on the support of the Bell
 states. The anticommuting observables are used to build the required local isometry circuits. We design the two-step swap isometry circuits for self testing the entanglement in spatial-mode and polarization DOFs, respectively. Our hyperentanglement self testing protocol has three advantages. First, it establishes a general hyperentanglement self testing framework.
 It can be extended to self test the multi-DOF hyperentanglement and the multi-particle hyperentanglement. Second, this hyperentanglement self testing framework can adopt the robust self testing protocols for one-DOF entanglement to construct the robust hyperentanglement self testing. Third, this hyperentanglement self-testing framework is feasible with current experimental technologies. It provides a simple and feasible self-testing method for complex hyperentanglement sources, and has potential application in future hyperentanglement-based quantum network.

 \textbf{Acknowledgment}
This work was supported by the National Natural Science Foundation of China under Grants No. 12175106, 92365110, and 12574393.

\end{document}